\documentstyle[11pt,epsfig]{article}
\newcommand{\bea}{\begin{eqnarray}}
\newcommand{\eea}{\end{eqnarray}}
\def\vel{\left|}
\def\ver{\right|}

\def\vel{\left|}
\def\ver{\right|}
\def\nnb{\nonumber}
\def\ga{\left(}
\def\dr{\right)}

\def\rar{\rightarrow}
\def\nnb{\nonumber}

\def\lla{\left<}
\def\rra{\right>}
\def\ba{\begin{array}}
\def\ea{\end{array}}

\begin{document}
\begin{titlepage}

\thispagestyle{empty}

\vspace{0.2cm}

\title{The sensitivity of the zero position of the forward--backward asymmetry
to new physics effects in the  $B \rar K^\ast \mu^+ \mu^-$ decay 
\author{Altu\u{g} Arda and M\"{u}ge Boz \\
Hacettepe University, Department of Physics,\\  
06532 Ankara, Turkey\\}}
\date{}
\maketitle

\begin{center}\begin{minipage}{5in}

\begin{center} ABSTRACT\end{center}
\baselineskip 0.2in
{Starting with the most general effective Hamiltonian 
comprising scalar and vector operators beyond the standard
model, we discuss the impact of various operators on the zero of 
the forward--backward asymmetry in the dileptonic $B$ decay 
$B \rar K^\ast \mu^+ \mu^-$. We find that, zero of the 
asymmetry is highly sensitive to the sign and size of the 
vector--vector operators and opposite chirality counterparts
of the usual operators. The scalar--scalar four--fermion
operators, on the other hand, have mild effect on the 
zero of the asymmetry. Our results are expected to be 
checked in the near future experiments.}
\end{minipage}
\end{center}
\end{titlepage}

\eject
\rm
\baselineskip=0.25in

\section{Introduction}

The flavour-changing transitions, which generally arise at one and higher
loop orders provide an excellent testing ground for the standart model (S.M). Moreover, 
it is with such decays that the new physics effects can be probed via the
loops of the particles beyond the S.M spectrum. Therefore, there are 
sound theoretical and experimental reasons for studying the 
flavour changing neutral current (FCNC)
processes. Among all the FCNC phenomena,  the rare $B$ decays \cite{munz} 
are especially  important, since one can both test the SM and search for possible NP effects, by confronting
the theoretical results  with the experiment. 

In addition to having 
already determined the branching ratio of $B\rightarrow 
X_s \gamma$ \cite{bsgam} and the CP asymmetry of 
$B\rightarrow J/\psi K$ \cite{acp}, experimental activity in $B$ physics, has
begun to probe FCNC phenomena in semileptonic $B$ decays 
\cite{belle,babar,cdf1,cleo1}, and 
these experiments are expected to give precise measurements 
in semileptonic decays in the near future. 

Concerning the semileptonic $B$ decays,  
$B \rar X_s \ell^+ \ell^-~
(X_s=K, K^{\ast},\  \ell=e,\mu,\tau)$ decay  is an example having both
theoretical and experimental importance.
The  forward--backward asymmetry $\mbox{A}_{\small \mbox{FB}}$
of these decays is a particularly interesting quantity, 
since it vanishes at a specific value of the dilepton invariant
mass \cite{burdman2,bsll} in a hadronically clean way. 
In the recent literature, the dilepton invariant mass spectra, and the forward backward asymmetry in
$B \rar X_s \ell^+ \ell^-~$ decays has been analyzed in the detailed work of \cite{bsll}, using  the 
large energy effective theory (LEET) approach,
and a simple analytic expression for the zero position of the
$\mbox{A}_{\small \mbox{FB}}$ in the S.M has been derived.
It has been found that 
the value of the dilepton invariant mass for which 
 $\mbox{A}_{\small \mbox{FB}}$ may become zero  provides a quite 
simple relation between the electric dipole coefficient $C_7$ and $C_9$,
which is nearly free of hadronic uncertainities \cite{bsll}.
Furthermore, the next to leading order (NLO) 
corrections to the exclusive decay has been  carried out in  \cite{karma1,karma2,ball,karma3}.

It is known that in the S.M, and in many of its extensions, the $B\rar  X_s \ell^+\ell^-$
decay is completely determined by the Wilson coefficients of only three
operators evaluated at the scale $\mu=m_b$.
On the other hand, the most general analysis of the  $B\rightarrow X_s \ell^+\ell^-$
decay, based on the  general four Fermi interaction, include new operators
beyond the usual set. The new structures in the effective Hamiltonian 
\cite{fukae1,aliev}
make these decays 
quite interesting as an alternative testing platform for the S.M, and
provide clues about the nature of the physics beyond the S.M.
In the literature, the  general model independent
analysis of the inclusive $b \rar s \ell^+ \ell^-$ decay, in terms of 10 types of local
four--Fermi interactions, has been performed  in Ref. \cite{fukae1},
and, a  systematic analysis 
of   the exclusive  $B \rar K^\ast \ell^+ \ell^-~$ decay has been  presented
in \cite{aliev}. Moreover,  a detailed study of the 
lepton  polarization asymmetries in $B\rightarrow X_s \ell^+\ell^-$
decay has been carried out in a rather general model in \cite{fukae2},
and, in the exclusive decay  $B \rar K^\ast \ell^+ \ell^-~$ in \cite{aliev2}.

In this work, based on the work of  \cite{aliev}, 
our aim is  to analyze the possible new physics effects stemming 
from the new structures in the effective Hamiltonian in the  $B \rar K^\ast \mu^+ \mu^-~$ decay. 
Among the several works on  $B \rar K^\ast \l^+ \l^-~$
decays in the existing literature, there are several papers, which discusses the subject
either  using the zero mass approximation (for instance, see \cite{kim}), 
or, by taking into account  the lepton mass effects \cite{aliev,korner2}. 
In our analysis, we include the lepton mass effects, and   
we study the influence of the new operators to 
the value of the dilepton invariant mass for which $\mbox{A}_{\small \mbox{FB}}$ vanishes,
by taking into account of its value in the S.M.
It will be seen that the position of the zero  shifts  
in accord with the new physics contributions to the Wilson coefficients.

The organization of the work is as follows. In Sec. 2, 
starting from the differential decay width of the exclusive  $B \rar K^\ast
\mu^+ \mu^-~$ decay, we calculate the numerator of the forward backward asymmetry 
whose intersectional value with the zero axes will determine the
zero-position of the forward backward asymmetry. In Sec. 3, we carry out the
numerical analysis to study  the 
dependence of the zero position of the  forward--backward asymmetry
on the new coefficients. We conclude in Sec. 4. 

\section{The Model}

The matrix element of the $B \rar K^\ast \ell^+ \ell^-$ decay at quark level
is described by $b \rar s \ell^+ \ell^-$ transition for which  the effective
Hamiltonian at $O(\mu)$ scale can be written as:
\bea
{\cal H}_{eff} &=& \frac{4 G_F}{\sqrt{2}} V_{tb}V_{ts}^\ast
\sum_{i=1}^{10} {\cal C}_i(\mu) \, {\cal O}_i(\mu)~,
\eea
where the full set of the operators ${\cal O}_i(\mu)$ and the corresponding
expressions for the Wilson coefficients ${\cal C}_i(\mu)$ in the SM 
are given in \cite{buras,misiak}.

Taking into account of the most general form of the effective Hamiltonian, 
there are ten independent local four--Fermi
interactions which may contribute to the process\cite{fukae1}.
Neglecting  the tensor type interactions
$C_{T}$ and $C_{TE}$ \cite{aliev}, the explicit form of the matrix element 
${\cal M}$ of the 
$b \rar s \ell^+ \ell^-$ transition can be written as a sum of the SM and 
new physics contributions, including eight independent local four--Fermi
interactions. 
\begin{eqnarray}
{\cal M} &=& 
\frac{G \alpha}{\sqrt{2} \pi} V_{tb}V_{ts}^\ast 
\Bigg\{ \ga  C_9^{eff} - C_{10} \dr \bar s_L \gamma_\mu b_L \,
\bar \ell_L \gamma^\mu \ell_L +
\ga C_9^{eff} + C_{10} \dr \bar s_L \gamma_\mu b_L \,
\bar \ell_R \gamma^\mu \ell_R \nnb \\
&-& 2 C_7^{eff} \bar s i \sigma_{\mu\nu}  \frac{\hat q^\nu}{\hat s} 
\ga \hat m_s L + \hat m_b R \dr b \, \bar \ell \gamma^\mu \ell+  
C_{LL} \bar s_L \gamma_\mu b_L \, \bar \ell_L
\gamma^\mu \ell_L+ 
C_{LR} \bar s_L \gamma_\mu b_L \, \bar \ell_R \gamma^\mu \ell_R\nnb\\
&+&
C_{RL} \bar s_R \gamma_\mu b_R \, \bar \ell_L \gamma^\mu \ell_L 
+ C_{RR} \bar s_R \gamma_\mu b_R \, \bar \ell_R \gamma^\mu \ell_R +
C_{LRLR} \bar s_L b_R \, \bar \ell_L  \ell_R 
+ C_{RLLR} \bar s_R b_L \, \bar \ell_L  \ell_R \nnb\\
&+&     
C_{LRRL} \bar s_L b_R \, \bar \ell_R  \ell_L 
+ C_{RLRL} \bar s_R b_L \, \bar \ell_R  \ell_L  \Bigg \}
\label{2}
\end{eqnarray}
Here,  $C_{LL(LR)}$,\,  $C_{RR(RL)}$,\,  $C_{LRLR(LRRL)}$,\,  $C_{RLLR(RLRL)}$,
are the coefficients of the four Fermi interactions. 
$R=(1+\gamma_5)/2$ and $L=(1-\gamma_5)/2$, and  
\bea
\hat s=q^2/m_B^2, ~~~~\hat m_b=m_b/m_B,~~~~q=(p_B-p_{K^\ast}) \nnb, 
\eea
The expression for $C^{eff}_9({\hat s})$  in Eq. (\ref{2}) is given by:
\begin{eqnarray}
C^{eff}_9({\hat s}) &=& C_9 +g( z,{\hat s})(3C_1+C_2+3C_3+C_4+3C_5+C_6)\nnb\\
&-&\frac{1}{2}g(1,{\hat s})(4C_3+4C_4+3C_5+C_6)\nnb\\
&-& \frac{1}{2}g(0,{\hat s})(C_3+3C_4)+\frac{2}{9}(3C_3+C_4+3C_5+C_6),~
\label{3}
\end{eqnarray}
where $ z=\frac{m_c}{m_b}$, the values of
$g( z,{\hat s})$, $g(1,{\hat s})$, $g(0,{\hat s})$
can be found in \cite{buras,misiak}, and the values of 
$C_i$ in the SM are given in the numerical analysis.

As mentioned in the Introduction, our aim is to find the value of the
dilepton invariant mass
for which $\mbox{A}_{\small \mbox{FB}}$ vanishes. Therefore,
in order to determine this quantity, following the work of  \cite{aliev}, we
first concentrate on  the
decay width  of  the $B \rar K^\ast \mu^+ \mu^-$  decay.
The matrix elements, $\lla K^\ast \vel \bar s \gamma_\mu (1 \pm \gamma_5) b
\ver B \rra$,  $\lla K^\ast \vel  \bar s i  \sigma_{\mu\nu} q^\nu (1 + \gamma_5) b \ver B \rra$,
$\lla K^\ast \vel \bar s (1 \pm \gamma_5) b \ver B \rra $
have been calculated  in \cite{aliev}. 
Using the matrix elements, and the helicity amplitude formalism \cite{ball,hagiwara,korner}, 
the decay width  of  the $B \rar K^\ast \mu^+ \mu^-$ 
decay is given by:
\bea
\frac{d\Gamma}{dq^2 du}&=& \frac{G^2 \alpha^2}{2^{14} \pi^5 m_B}
\vel V_{tb} V_{ts}^\ast \ver^2 
v \lambda^{1/2}(1,r,s)  \sum_{i=1}^6  {\cal M}_{i} ^2,
\label{decaywidth}
\eea
where, $v$ is the velocity of $\mu$, and
\bea
&&v = \sqrt{1 - 4 {\hat m}_\mu^2/ \hat s},~~~{\hat m}_\mu=\frac{m_\mu}{m_B},~\nnb\\
&&\lambda(1,{\hat m}_K,\hat s) = 1 +{\hat m }^4_{K^{\ast}} + \hat s^2 - 2  {\hat m }^2_{K^{\ast}}  
\hat s - 2 {\hat m }^2_{K^{\ast}} - 2 \hat s~, \nnb \\
&&{\hat m }^2_{K^{\ast}}=m_{K^*}^2/m_B^2~,
\label{4}
\eea
Here,  ${\cal M}_{i}^2$ are  the   
the combinations of the matrix elements which can be written in the following form:
\bea 
 {\cal M}_{1}^2&=& \vel {\cal M}^{++}_{+}\ver^2+\vel {\cal
M}^{++}_{-}\ver^2~,\nnb\\ 
{\cal M}_{2}^2&=& \vel {\cal M}^{--}_{+}\ver^2+\vel {\cal
M}^{--}_{-}\ver^2~,\nnb\\ 
{\cal M}_{3}^2&=&\vel {\cal M}^{+-}_{+}\ver^2+ \vel {\cal
M}^{+-}_{-}\ver^2~,\nnb\\ 
 {\cal M}_{4}^2&=&\vel {\cal M}^{-+}_{+}\ver^2+ \vel {\cal
M}^{-+}_{-}\ver^2~,\nnb\\ 
{\cal M}_{5}^2&=&\vel {\cal M}^{+-}_{0}\ver^2+ \vel {\cal
M}^{-+}_{0}\ver^2~,\nnb\\ 
 {\cal M}_{6}^2&=& \vel {\cal M}^{++}_{0}\ver^2+\vel {\cal
M}^{--}_{0}\ver^2~,
\label{5}
\eea
where, the  superscripts correspond to the helicity of the  $K^\ast$
meson, and the subscripts denote  the helicities of the  muons.
The combinations of the matrix elements $ {\cal M}_{i}^2$
include the Wilson coefficients $C_7$, $C_9$, $C_{10}$
and the new operators beyond the usual set. Therefore, for
convenience, we define: 
\bea
\label{6}
{\tilde{C}}_{7}&=&4 C_{eff}~ \nnb,
\eea
and, the combinations of the new coefficients as:
\bea
\label{7}
{\tilde{C}}_{RR_{+}}&=&C_{RR} + C_{RL}~,\nnb\\
{\tilde{C}}_{RR_{-}}&=&C_{RR} - C_{RL}~,\nnb\\
{\tilde{C}}_{LL_{+}}&=&C_{LL} + C_{RL}~,\nnb\\
{\tilde{C}}_{LRLR_{-}}&=&C_{LRLR} - C_{RLLR}~,\nnb\\
{\tilde{C}}_{LRRL_{-}}&=&C_{LRRL} - C_{RLRL}~,\nnb\\
{\tilde{C}}_{9}&=&2 C_{9}^{eff} +  C_{LL}+C_{LR}~,\nnb\\
{\tilde{C}}_{10}&=&2 C_{10}^{eff} -C_{LL}+C_{LR}~,
\eea
Here, the coefficients ${\tilde{C}}_{9}$, and ${\tilde{C}}_{10}$
describe the contributions from the S.M, and the new physics.
The combinations  ${\tilde{C}}_{LRLR_{-}}$, and ${\tilde{C}}_{LRRL_{-}}$
desribe scalar type interactions.
Then, using the explicit forms of the matrix elements \cite{aliev}, 
we calculate $ {\cal M}_{1}^2$, and $ {\cal M}_{2} ^2$  as:
\bea
{\cal M}_{1}^2&=& 2 m_\mu^2 (1-u^2)\ f_1~,\nnb\\
 {\cal M}_{2} ^2&=& {\cal M}_{2}^2~,
\label{8}
\eea
where,
\bea
u=\cos\theta,
\label{9}
\eea
and  $\theta$ is the angle between $K^\ast~$, and 
$\mu^-$. Here,  $f_1$ read as: 
\bea
f_1&=& \big(H_{+}^2+ H_{-}^2 \big)
\vel {\tilde{C}}_{9}\ver^2
+\big( \frac {m_b}{\hat s m_B^2}\big)
\big(H_{+}{\cal H}_{+}+ H_{-}{\cal H}_{-}\big)
2 Re [{\tilde{C}}_{9}{\tilde{C}}_{7}^{*}]\nnb\\
&+&\big(H_{+}h_{+}+ H_{-} h_{-}\big)
2 Re [{\tilde{C}}_{9}{\tilde{C}}_{RR_{+}}^{*}]
+\big (\frac{m_b}{\hat s m_B^2}\big)
\big( h_{+}{\cal H}_{+}+ h_{-}{\cal H}_{-}\big)
2 Re [{\tilde{C}}_{7}{\tilde{C}}_{RR_{+}}^{*}]\nnb\\
&+&\big( \frac {m_b}{\hat s m_B^2}\big)^2
\big({\cal H}_{+}^2+ {\cal H}_{-}^2\big)
 \vel {\tilde{C}}_{7}\ver^2
+\big( h_{+}^2+ h_{-}^2\big)
\vel {\tilde{C}}_{RR_{+}} \ver^2~,
\label{10}
\eea
The expressions of  $ {\cal M}_{3}^2$
and  ${\cal M}_{4}^2$ 
are given by:
\bea
 {\cal M}_{3}^2&=& \frac{ \hat s m_B^2 }{2} \bigg \{ (1+u^2)f_3^{(+)}-2 u
f_3^{(-)}\bigg\}~,  
\label{11}
\eea
\bea
{\cal M}_{4}^2&=& \frac{ \hat s m_B^2}{2}\bigg\{
(1+u^2)f_{3}^{ (+)}(v\rightarrow -v)+2 u
f_{3}^{(-)} (v \rightarrow -v)\bigg\}~,  
\label{12}
\eea
where
\bea
f_{3}^{(\pm)}&=&  
\big( \frac {m_b}{\hat s m_B^2}\big)
\big(H_{+}{\cal H}_{+}\pm H_{-}{\cal H}_{-}\big)
\big (2 Re [{\tilde{C}}_{9}{\tilde{C}}_{7}^{*}]
+v 2 Re [{\tilde{C}}_{10}{\tilde{C}}_{7}^{*}]\big)\nnb\\
&+&\big(H_{+}h_{+}\pm H_{-} h_{-}\big)
\big(2 Re [{\tilde{C}}_{9}{\tilde{C}}_{RR_{+}}^{*}]
+v\ 2 Re [{\tilde{C}}_{9}{\tilde{C}}_{RR_{-}}^{*}]\nnb\\
&+& v \ 2  Re [{\tilde{C}}_{10}{\tilde{C}}_{RR_{+}}^{*}]
+ v^2\ 2 Re [{\tilde{C}}_{10}{\tilde{C}}_{RR_{-}}^{*}]\big)\nnb\\
&+&
\big (\frac{m_b}{\hat s m_B^2}\big)
\big(h_{+}{\cal H}_{+}\pm h_{-}{\cal H}_{-}\big)
\big(2 Re [{\tilde{C}}_{7}{\tilde{C}}_{RR_{+}}^{*}]
+ v \ 2 Re [{\tilde{C}}_{7}{\tilde{C}}_{RR_{-}}^{*}]\big)\nnb\\
&+&\big( \frac {m_b}{\hat s m_B^2}\big)^2
\big({\cal H}_{+}^2 \pm {\cal H}_{-}^2\big)
  \vel {\tilde{C}}_{7}\ver^2\nnb\\
 &+& \big( H_{+}^2\pm H_{-}^2 \big)
\big (  \vel {\tilde{C}}_{9}\ver^2
+v\ 2 Re [{\tilde{C}}_{9}{\tilde{C}}_{10}^{*}]
+ v^2 \vel {\tilde{C}}_{10}\ver^2
\big)\nnb\\
&+&\big( h_{+}^2\pm h_{-}^2\big)
\big(\vel {\tilde{C}}_{RR_{+}}\ver^2
+v\ 2 Re [{\tilde{C}}_{RR_{+}}{\tilde{C}}_{RR_{-}}^{*}]
+v^2 \vel {\tilde{C}}_{RR_{-}}\ver^2 \big)~, 
\label{13}
\eea
In Eq. (\ref{10}), and Eq. (\ref{13}) the functions
$H_{+}$, $h_{-}$  are the helicity amplitudes, and 
in terms of the form factors, they have the following structures:  
\bea
H_\pm &=& m_B \Bigg[ \pm \lambda^{1/2} 
\frac{V(\hat s )}{1+{\hat m}_K} + (1+{\hat m}_K) A_1(\hat s ) \Bigg]~, \nnb \\ 
{\cal H}_\pm &=& 2 m_B^2\left[ \pm \lambda^{1/2} 
T_1(\hat s ) +
(1-{\hat m}^2_K)T_2(\hat s ) \right]~, \nnb \\ 
h_{\pm}&=&H_{\pm} (A_1 \rightarrow -A_1,\,\,\, A_2 \rightarrow -A_2)   
\label{14}
\eea
Finally, ${\cal M}_{5}^2$, and ${\cal M}_{6} ^2$
can be calculated as:
\bea
 {\cal M}_{5}^2&=& 2 \ \hat s m_B^2 \ (1-u^2) \ f_5~,  
\label{15}
\eea
where
\bea
f_5&=&
H_0^2\big (  \vel {\tilde{C}}_{9}\ver^2
+v^2 \  \vel {\tilde{C}}_{10}\ver^2\big)
-\big( \frac {m_b}{\hat s m_B^2}\big) H_{0}{\cal H}_{0}\
\big( 2 Re [{\tilde{C}}_{9}{\tilde{C}}_{7}^{*}]\big) \nnb\\
&+&H_0 \ h_0
\big(2 Re [{\tilde{C}}_{9}{\tilde{C}}_{RR_{+}}^{*}]
+v^2 \ 2 Re [{\tilde{C}}_{10}{\tilde{C}}_{RR_{-}}^{*}]\big)
+\big( \frac {m_b}{\hat s m_B^2}\big)^2 {\cal H}_{0}^2
\vel {\tilde{C}}_{7}\ver^2\nnb\\
&-&\big( \frac {m_b}{\hat s m_B^2}\big)
{\cal H}_{0} h_{0}2 Re [{\tilde{C}}_{7}{\tilde{C}}_{RR_{+}}^{*}]+
h_0^2\big(\vel {\tilde{C}}_{RR_{+}}\ver^2
+v^2 \vel {\tilde{C}}_{RR_{-}}\ver^2\big)~, 
\label{16}
\eea
and,
\bea
{\cal M}_{6}^2&=&4 m_\mu^2 u^2 f_6^{(1)}  
+ 4 m_\mu u  f_6^{(2)}+ 4 m_\mu^2 f_6^{(3)}~,  
\label{17}
\eea
with
\bea
f_6^{(1)}&=&
2 (H_{0})^2
 \vel {\tilde{C}}_{9}\ver^2  
- 2\big( \frac{m_b}{\hat s m_B^2}\big)
H_{0} {\cal H}_{0}  
\big (2 Re [{\tilde{C}}_{9}{\tilde{C}}_{7}^{*}]\big)\nnb\\ 
&+& H_{0}h_{0} 
\big(2 Re [{\tilde{C}}_{9}{\tilde{C}}_{LL_{+}}^{*}]
+ 2 Re [{\tilde{C}}_{9}{\tilde{C}}_{RR_{+}}^{*}]\big)\nnb\\ 
&+& 
2 \big( \frac{m_b}{\hat s m_B^2}\big)^2  {\cal H}_{0}^2 
\vel {\tilde{C}}_{7}\ver^2-
 \big (\frac{m_b}{\hat s m_B^2}\big) h_{0} {\cal H}_{0}  
\big(2 Re [{\tilde{C}}_{7}{\tilde{C}}_{LL_{+}}^{*}]
+\ 2 Re [{\tilde{C}}_{7}{\tilde{C}}_{RR_{+}}^{*}]\big) \nnb\\
&+&
h_0^2\big(\vel {\tilde{C}}_{LL_{+}}\ver^2+ \vel
{\tilde{C}}_{RR_{+}}\ver^2\big)~,
\label{18}
\eea
\bea
 f_6^{(2)} &=&  
m_\mu H_s^0 h_0
\big(2 Re [  {\tilde{C}}_{RR_{+}}{\tilde{C}}_{10} ^{*}]
- 2 Re [  {\tilde{C}}_{LL_{+}}{\tilde{C}}_{10} ^{*}]
\big)\nnb\\
&+&\big( \frac {s m_B^2}{ m_b}\big)
H_s^0 h_0
\Bigg \{
(1-v)\Big( 2 Re [{\tilde{C}}_{LL_{+}}{\tilde{C}}_{LRRL_{-}}^{*}]
- 2 Re [{\tilde{C}}_{RR_{+}}{\tilde{C}}_{LRLR_{-}}^{*}]\Big)\nnb\\
&-&(1+v)\Big( 2 Re [{\tilde{C}}_{LL_{+}}{\tilde{C}}_{LRLR_{-}}^{*}]
+ 2 Re [{\tilde{C}}_{RR_{+}}{\tilde{C}}_{LRRL_{-}}^{*}]\Big)
\Bigg\} \nnb\\
&+&h_0 h_s^0 m_\mu
\big(2 Re [{\tilde{C}}_{RR_{+}}{\tilde{C}}_{RR_{-}}^{*}]
-2 Re [{\tilde{C}}_{LL_{+}}{\tilde{C}}_{RR_{-}}^{*}]\big)\nnb\\
&-& 2\big( \frac{\hat s m_B^2}{m_b}\big) 
H_{0} H_s^0 v
\big( 2 Re [{\tilde{C}}_{9}{\tilde{C}}_{LRLR_{-}}^{*}]
+2 Re [{\tilde{C}}_{9}{\tilde{C}}_{LRRL_{-}}^{*}]\big)\nnb\\
&+&2 H^{0}_{s} {\cal H}_{0}
v \big(2 Re [{\tilde{C}}_{7}{\tilde{C}}_{LRLR_{-}}^{*}]
+2 Re [{\tilde{C}}_{7}{\tilde{C}}_{LRRL_{-}}^{*}]\big)~,
\label{19}
\eea
\bea
f_6^{(3)}&=& 
2 (h_s^0)^2
\vel {\tilde{C}}_{RR_{-}}\ver^2
+2 h_s^0\ H_s^0
\big( 2 Re [{\tilde{C}}_{10}{\tilde{C}}_{RR_{-}}^{*}]\big) \nnb\\
&+& 2 \big( \frac{\hat s}{{\hat m}_\mu {\hat m}_b}\big) h_s^0\ H_s^0
\big( 2 Re [{\tilde{C}}_{RR_{-}}{\tilde{C}}_{LRLR_{-}}^{*}]
- 2 Re [{\tilde{C}}_{RR_{-}}{\tilde{C}}_{LRRL_{-}}^{*}]\big)\nnb\\
&+& ( H_s^0)^2  \Big \{
\vel { 2 \tilde{C}}_{10} \ver^2
+2 \big( \frac{\hat s}{{\hat m}_\mu {\hat m}_b}\big)
\big( 2 Re [{\tilde{C}}_{10}{\tilde{C}}_{LRLR_{-}}^{*}]
-2 Re [{\tilde{C}}_{10}{\tilde{C}}_{LRRL_{-}}^{*}]\big)\nnb\\
&+&2 \big (\frac{\hat s}{{\hat m}_\mu {\hat m}_b}\big)^2
\big\{ (1+v^2)\big(
\vel {\tilde{C}}_{LRLR_{-}}\ver^2
+\vel {\tilde{C}}_{LRRL_{-}}\ver^2\big)\nnb\\
&-& (1-v^2)  2 Re [{\tilde{C}}_{LRLR_{-}}{\tilde{C}}_{LRRL_{-}}^{*}]\big\}
\Big\}~,
\label{20}
\eea
where, 
\bea
H_0 &=& \frac{m_B}{2 {\hat m}_K \sqrt { \hat s}} \Bigg[
- (1-{\hat m}^2_K-\hat s) (1+{\hat m}_K) A_1(\hat s) +
 \lambda  \frac{A_2 (\hat s)}{1+{\hat m}_K}
\Bigg]~, \nnb \\    
H_S^0 &=& - \frac{m_B \lambda^{1/2}}{ \sqrt{\hat s} } A_0 (\hat s ) \nnb \\ 
{\cal H}_\pm &=& 2 m_B^2\left[ \pm \lambda^{1/2} 
T_1(\hat s ) +
(1-{\hat m}^2_K)T_2(\hat s ) \right]~, \nnb \\ 
{\cal H}_0 &=& \frac{m_B^2}{ {\hat m}_{K} \sqrt{\hat s } }
\Bigg\{ (1- {\hat m}^2_K- \hat s) (1- {\hat m}^2_K) T_2(\hat s )
-\lambda  
\left[ T_2(\hat s )
+ \frac{\hat s}{1-{\hat m}^2_K} T_3(\hat s ) \right] \Bigg\}~,\nnb\\
h_0&=&H_0 (A_1 \rightarrow -A_1,\,\,\, A_2 \rightarrow -A_2)~,   
\label{21}
\eea
As mentioned in the Introduction,
our aim is to determine the zero position of the forward backward asymmetry, 
\bea
\frac{d}{dq^2}A_{FB}(q^2) = \frac{\displaystyle{
\int_0^1dx \frac{d\Gamma}{dq^2 dx} - \int_{-1}^0dx \frac{d\Gamma}{dq^2 dx}}}
{\displaystyle{
\int_0^1dx \frac{d\Gamma}{dq^2 dx} + \int_{-1}^0dx \frac{d\Gamma}{dq^2 dx}}}~,
\label{22}
\eea
which can predict possibly the new physics contributions. Indeed,
existence of the new physics can be confirmed by the shift in the
zero position of the forward backward asymmetry\cite{bsll}.
Therefore, using Eqs.(\ref{8})-(\ref{20}), we calculate the the numerator of the forward backward asymmetry
as: 
\bea
{\cal N}&=& \frac{G^2 \alpha^2}{2^{14} \pi^5} \vel V_{tb} V_{ts}^\ast\ver^2
{\cal R}~,
\label{23}
\eea
where
\bea
{\cal R}&=& 
\frac{1}{m_B} v \lambda^{1/2}
\Bigg[
2 \hat s m_B^2 v \Bigg \{\big(h_{-}^2-h_{+}^2\big)\
2 Re[{\tilde{C}}_{RR_{+}}\ {\tilde{C}}_{RR_{-}}^{*}]
+  \big(  H_{-}^2-  H_{+}^2 \big)
\ 2 Re[{\tilde{C}}_{9}\ {\tilde{C}}_{10}^{*}]\nnb\\
&+& \big(H_{-}h_{-}-H_{+}h_{+}\big)
\big(2 Re[ {\tilde{C}}_{9}\ {\tilde{C}}_{RR_{-}}^{*}]+
2 Re[{\tilde{C}}_{10}\ {\tilde{C}}_{RR_{+}}^{*}]\big)\nnb\\
&+&\big( \frac{m_b}{\hat s m_B^2}\big) \Bigg( \big({\cal H}_{-} h_{-}-{\cal
H}_{+} h_{+}\big) 2  Re [{\tilde C}_7\  {\tilde{C}}_{RR_{-}}^{*}] 
+ \big( {\cal H}_{-} H_{-} -{\cal
H}_{+}  H_{+} \big) 2 Re [{\tilde C}_7 \ {\tilde{C}}_{10}^{*}]\Bigg)\Bigg\}\nnb\\
&+&4 m_\mu^2 \Bigg\{
h_{0}\ h_s^{0} 
\big( 2 Re[ {\tilde{C}}_{RR_{+}}{\tilde{C}}_{RR_{-}}^{*}]
-2Re[{\tilde{C}}_{LL_{+}}{\tilde{C}}_{RR_{-}}^{*}]\big)\nnb\\
&+& 
h_{0}\ H_s^{0}\big( 2 Re[{\tilde{C}}_{RR_{+}}{\tilde{C}}_{10}^{*}]
-2Re[ {\tilde{C}}_{LL_{+}}{\tilde{C}}_{10}^{*}]\big)\Bigg\}\nnb\\
&+&
\frac{8 m_\mu \hat s m_B^2 v}{m_b} 
\Bigg\{
 \frac{m_b}{\hat s m_B^2}  {\cal H}_{0}\ H_s^{0}
 \Big( 2 Re [{\tilde C}_7  {\tilde{C}}_{LRLR_{-}}^{*}]
+2 Re [{\tilde C}_7  {\tilde{C}}_{LRRL_{-}}^{*}]\Big)\nnb\\
&-& H_{0}\ H_s^{0}   
\Big(2 Re [{\tilde{C}}_{9} {\tilde{C}}_{LRLR_{-}}^{*}]
+2 Re [{\tilde{C}}_{9} {\tilde{C}}_{LRRL_{-}}^{*}]\Big)\Bigg\}\nnb\\
&+&
\frac{4 m_\mu \hat s m_B^2}{m_b} 
 h_{0}\ H_s^{0} 
\Bigg \{
(1-v)\Big( 2 Re [{\tilde{C}}_{RR_{+}}{\tilde{C}}_{LRLR_{-}}^{*}]
+ 2 Re [{\tilde{C}}_{LL_{+}}{\tilde{C}}_{LRRL_{-}}^{*}]\Big)\nnb\\
&-&(1+v)\Big( 2 Re [{\tilde{C}}_{RR_{+}}{\tilde{C}}_{LRRL_{-}}^{*}]
+ 2 Re [{\tilde{C}}_{LL_{+}}{\tilde{C}}_{LRLR_{-}}^{*}]\Big)
\Bigg\} \Bigg]~,
\label{24}
\eea
Then, using  Eq.~(\ref{14}), and  Eq.~(\ref{21}), ${\cal R}$ can be written 
in the following form:
\bea
{\cal R}&=&\frac{1}{m_B} v \lambda^{1/2}\Bigg\{ 
{\cal R}_{1}\hat {s} v \lambda^{1/2} A_1 V-
{\cal R}_{2}v\lambda^{1/2} T_2 V +
{\cal R}_{3} v
\lambda^{1/2} A_1 T_1\nnb\\
&-&
{\cal R}_{4} \lambda^{1/2} A_0 A_1 (1- {\hat m }^2_{K^{\ast}}-\hat{s})
(\frac{{\cal R}_{5}}{\hat s}+{\cal R}_{6}v+{\cal R}_{7})
+{\cal R}_{8} \lambda^{3/2} A_0 A_2 (\frac{ {\cal R}_{5}}{ \hat{s}}+{\cal R}_{6}
v+{\cal R}_{7})\nnb\\ 
&-&{\cal R}_{9} \ v \lambda^{1/2}
A_0 T_2 (1- 3 {\hat m }^2_{K^{\ast}}-\hat{s})+ {\cal R}_{10} v \lambda^{3/2}
A_0 T_3  \Bigg\}~,
\label{25}
\eea
Here,
\bea
{\cal R}_{1}&=&8\ m_B^4\Big(2 Re[{\tilde{C}}_{RR_{+}}\
{\tilde{C}}_{RR_{-}}^{*}]-2Re[{\tilde{C}}_{9}
\ {\tilde{C}}_{10}^{*}]\Big)~,\nnb\\
{\cal R}_{2}&=&8 m_B^4 \ {\hat m}_{b}\  
( 1-{\hat m}_{K}) \Big( 2 Re [ {\tilde C}_7 {\tilde{C}}_{10}^{*}]+2
Re[{\tilde C}_7
{\tilde{C}}_{RR_{-}}^{*}]\Big)~,\nnb\\ 
{\cal R}_{3}&=&8 \ m_B^4 \ {\hat m}_{b} \ 
( 1+ {\hat m }_{K^{\ast}}) \Big(-2 Re [{\tilde C}_7 {\tilde{C}}_{10}^{*}]+
2 Re [{\tilde C}_7 {\tilde{C}}_{RR_{-}}^{*}]\Big)~,\nnb\\ 
{\cal R}_{4}&=& \frac{ 4 m_B^4} { {\hat m }_{K^{\ast}}}( 1+   {\hat m }_{K^{\ast}} )~,\nnb\\ 
{\cal R}_{5}&=& \frac{{{\hat m}}_{\mu}^{2}}{2}
\Big( 2 Re[{\tilde{C}}_{RR_{+}}{\tilde{C}}_{10}^{*}] 
+2 Re [{\tilde{C}}_{RR_{+}}{\tilde{C}}_{RR_{-}}^{*}]
-2 Re [{\tilde{C}}_{LL_{+}}{\tilde{C}}_{10}^{*}]
-2 Re [{\tilde{C}}_{LL_{+}}{\tilde{C}}_{RR_{-}}^{*}]\Big)~,\nnb\\
{\cal R}_{6}&=& \frac { {{\hat m}}_{\mu}}{2 {{\hat m}}_{b}}  
\Big(4 Re [{\tilde{C}}_{9} {\tilde{C}}_{LRLR_{-}}^{*}]
+4 Re [{\tilde{C}}_{9} {\tilde{C}}_{LRRL_{-}}^{*}]
- 2 Re [{\tilde{C}}_{LL_{+}}{\tilde{C}}_{LRRL_{-}}^{*}]\nnb\\
&-& 2 Re [{\tilde{C}}_{LL_{+}}{\tilde{C}}_{LRLR_{-}}^{*}]
-2 Re [{\tilde{C}}_{RR_{+}}{\tilde{C}}_{LRRL_{-}}^{*}]
-2 Re [{\tilde{C}}_{RR_{+}}{\tilde{C}}_{LRLR_{-}}^{*}]\Big)~,\nnb\\
{\cal R}_{7}&=& \frac { {{\hat m}}_{\mu}}  { 2 {{\hat m}}_{b}}  
\Big( 2 Re [{\tilde{C}}_{RR_{+}}{\tilde{C}}_{LRLR_{-}}^{*}]
+ 2 Re [{\tilde{C}}_{LL_{+}}{\tilde{C}}_{LRRL_{-}}^{*}]\nnb\\
&-&  2 Re [{\tilde{C}}_{RR_{+}}{\tilde{C}}_{LRRL_{-}}^{*}]
- 2 Re [{\tilde{C}}_{LL_{+}}{\tilde{C}}_{LRLR_{-}}^{*}]\Big)~,\nnb\\
{\cal R}_{8}&=& \frac{4 m_B^4}  {  {\hat m }_{K^{\ast}} (1+ {\hat m }_{K^{\ast}} ) }~,\nnb\\
{\cal R}_{9}&=& 8 m_B^4 \  \frac { {{\hat m}}_{\mu}}  {{\hat m }_{K^{\ast}}
} \Big(
2 Re [{\tilde C}_7 {\tilde{C}}_{LRLR_{-}}^{*}]+ 2 Re[{\tilde C}_7
{\tilde{C}}_{LRRL_{-}}^{*}]\Big)~,\nnb\\
{\cal R}_{10}&=& 8 m_B^4 \   \frac { { {\hat m}}_{\mu}}  
{ {\hat m }_{K^{\ast}} (1- {\hat m }^2_{K^{\ast}  }) } \Big(
2  Re [{\tilde C}_{7}  {\tilde{C}}_{LRLR_{-}}^{*}]+ 2 Re [\bar C_7
{\tilde{C}}_{LRRL_{-}}^{*}] \Big)~.
\label{26}
\eea
Naturally,  to find the zero position of  $\mbox{A}_{\small \mbox{FB}}$,
it is reasonable to find the roots of the function
${\cal R}$. However, due to the effective coefficient ${\tilde{C}}_{9}$,
the computation is quite complicated. Therefore,
in determining the zero position of  $\mbox{A}_{\small \mbox{FB}}$,
we analyze the variation of 
function ${\cal R}$ with the dilepton invariant mass.
The  intersectional value of ${\cal  R}$ with the zero axes will
determine the zero position of $\mbox{A}_{\small \mbox{FB}}$,  which can  
be interesting as an alternative 
testing platform  for the S.M, and provide clues about nature of the new
operators beyond the S.M. 

\section{Numerical analysis}

In the following we will perform a numerical analysis, to study the sensitivity of the zero position of 
the forward backward asymmetry to new physics effects, and discuss its
phenomenological implications. 
The zero position of the  $\mbox{A}_{\small \mbox{FB}}$ has been calculated in the S.M \cite{bsll},
\bea
Re[C_9^{eff}(\hat s)]=-\frac{{\hat m}_b}{\hat s} C_7^{eff}\Big \{ \frac{T_2(\hat
s)}{A_1(\hat s )} (1- \hat m_{K^{\ast}})+
\frac{T_1(\hat
s)}{V(\hat s )} (1+ \hat m_{K^{\ast}}) \Big \}~,
\label{27}
\eea
which depends on the ratio of the form factors, as well as the other
quantities. Thus, in principle the expression Eq.(\ref{27})
is affected by the presence of the hadronic form factors
making it a more uncertain relation.
However, using the large energy expansion theory (LEET),
it has been shown in \cite{bsll} that
both ratios of the form factors have no hadronic uncertainity
since the dependence of the intrinsically non-perturbative
quantities cancels, and the position of zero
in $B\rightarrow K^{\ast} \ell^{+}\ell^-$ is predicted simply in terms
of the short distance Wilson coefficients $C_9^{eff}$
and $C_7^{eff}$.
\bea
Re[C_9^{eff}(\hat s)]=- 2 \frac{ {\hat m}_b}{\hat s} C_7^{eff}
\frac{1-\hat s}{1+\hat m^2_{K^{\ast}}-\hat s}~,
\label{28}
\eea
Therefore, they have shown in \cite{bsll}  that the form factor dependence in  $ \hat s $ 
cancels in the  the large energy expansion approximation.
Moreover, they have found  the value of the zero position as 
${\hat s}=0.1 \mbox{GeV}$, with the numerical values of the coefficients at $\mu = m_{b}$
within the S.M,
\bea
\label{29}
C_1 &=&-0.248,~~~C_2=1.107,~~~C_3=0.011, \nnb \\
C_4&=& -0.026,~~~C_5=0.007,~~~C_6=-0.031,\nnb\\
C_7&=& -0.313,~~~C_9=4.344,~~~C_{10}=-4.669\nnb,~
\eea
and for $m_b = 4.4~\mbox{GeV}$.
To find  a reasonable agreement with the results of S.M,
we use the same input parameters for  $m_{b}$, and $C_i$, in our analysis.
However, we choose light cone QCD sum rules
method predictions for the form factors \cite{ball}.
Thus, using the results of \cite{ball}, in which the form factors are described 
by a three parameter fit, the ${\hat s}$ dependence of any of the  form factors  
appearing in Eq. (\ref{14}) could be parametrized as:
\begin{eqnarray}
\label{30}
F(\hat s) = \frac{F(0)}{1-a_F\, \hat s + b_F\, {\hat s}^{2}}~, 
\end{eqnarray}
The parameters for $F_0$, $a_F$, and $b_F$ 
for each form factor are given by: 
\bea
\label{31}
A_{0}&=&0.47,~~~a_F=1.64,~~~b_F=~0.94, \nnb \\
A_{1}&=&0.35,~~~a_F=0.54,~~~b_F=-0.02,\nnb\\
A_{2}&=&0.30,~~~a_F=1.02,~~~b_F=~0.08\nnb\\
V_{1}&=&0.47,~~~a_F=1.50,~~~b_F=~0.51, \nnb\\
T_{1}&=&0.19,~~~a_F=1.53,~~~b_F=~1.77, \nnb\\
T_{2}&=&0.19,~~~a_F=0.36,~~~b_F=-0.49, \nnb\\
T_{2}&=&0.13,~~~a_F=1.07,~~~b_F=~0.16 \nnb.
\eea
Moreover, in  the  numerical analysis,
we  use the  the kinematical range for the normalized dilepton invariant mass in terms of
the lepton and pseudo scalar masses: 
\bea
4 m_{\ell}^{2}/m_{B}^{2} \leq \hat s \leq (1-m_K^\ast/m_{B})^{2}.
\label{32}
\eea
In what follows, we  will analyze the variation of function ${\cal R}$
with the dilepton invariant mass.
In forming the scatter plots, we first consider
the case where all the new coefficients are zero
to investigate whether one can find a reasonable agreement with the results of
the S.M, and then let $C_{10}$, and $-C_{10}$ values for each of the new
coefficients,   setting 
all the others zero, to analyze the shift in
the zero position of  ${\cal R}$, as compared to the S.M.  
We would like to note that, in each scatter plot, the function ${\cal R}$
is divided by $10^3$, for convenience.
\begin{figure}[htb]
\vskip -3.0truein
\centering
\epsfxsize=6in
\leavevmode\epsffile{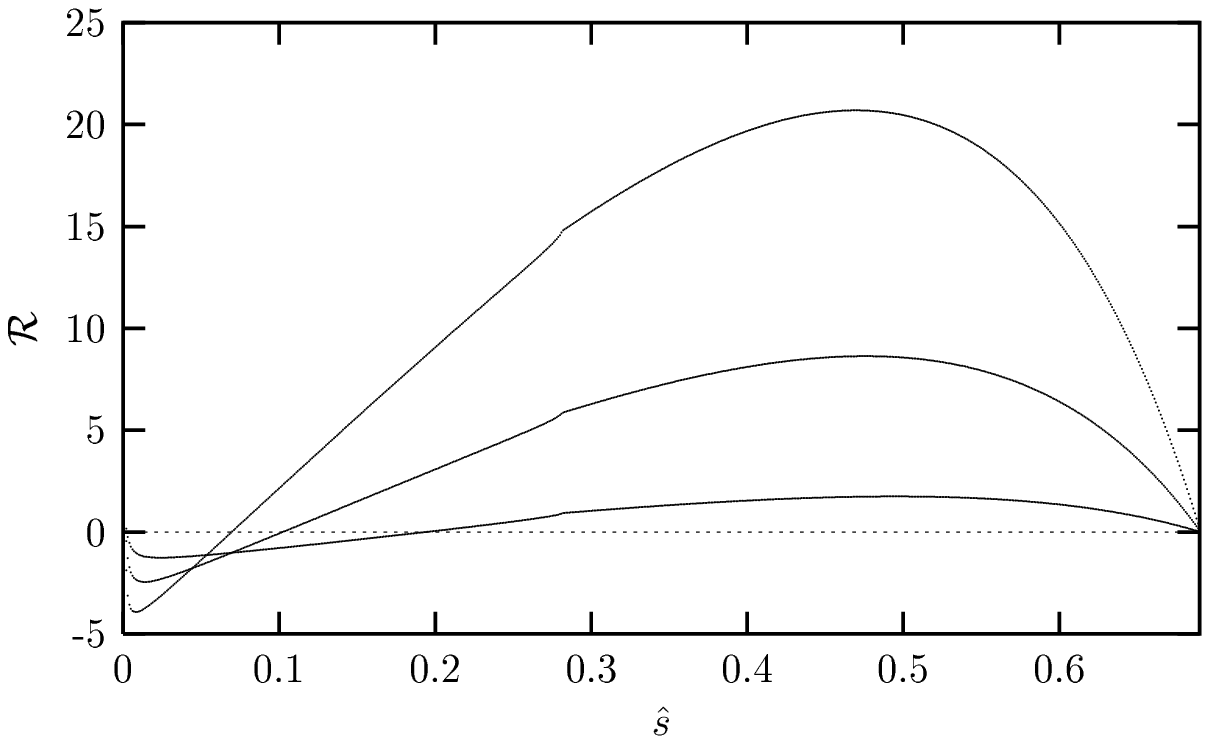}
\vskip -3.3truein
\caption[]{The dependence of ${\cal  R}$ on ${\hat s}$ 
which corresponds to the cases : $C_{LL}=-C_{10}$ (top curve),
$C_{LL}=C_{10}$ (bottom 
curve), and all the new coefficients are zero (middle curve) .}
\label{fig1}
\end{figure}

In Fig. 1, we show the dependence of ${\cal  R}$ on ${\hat s}$
when  $C_{LL}=-C_{10}$ for the top curve,  when $C_{LL}=C_{10}$ for the bottom curve,
and when all the new coefficients are set to zero for the middle curve
(as mentioned before, in plotting the top and the bottom curves,
we have set all the other new coefficients to zero).
As we can see from the figure that,
when all the new coefficients are
set to zero, the zero position of the forward-backward asymmetry occurs
at $\hat{s}\sim 0.1$, which is consistent with the results of  S.M  \cite{bsll}.
Naturally, for non-zero values of the new operators,
we expect the zero position of  ${\cal R}$ to shift from its value of S.M.
Indeed, when $C_{LL}=-C_{10}$
and for all the other coefficients
set equal to zero, ${\cal R}$ crosses zero around ${\hat s} \sim 0.06$. 
On the other hand,  when  $C_{LL}=C_{10}$,
and  for all the other coefficients
set equal to zero, the position of zero gradually shifts to $\sim 0.2$.  
A closer comparative look at the figure suggests that when $C_{LL}=-C_{10}$,
the position of the zero shifts to left, and when  $C_{LL}=C_{10}$,
it shifts to right, as compared to its value of S.M.
\begin{figure}[htb]
\vskip -3.0truein
\centering
\epsfxsize=6in
\leavevmode\epsffile{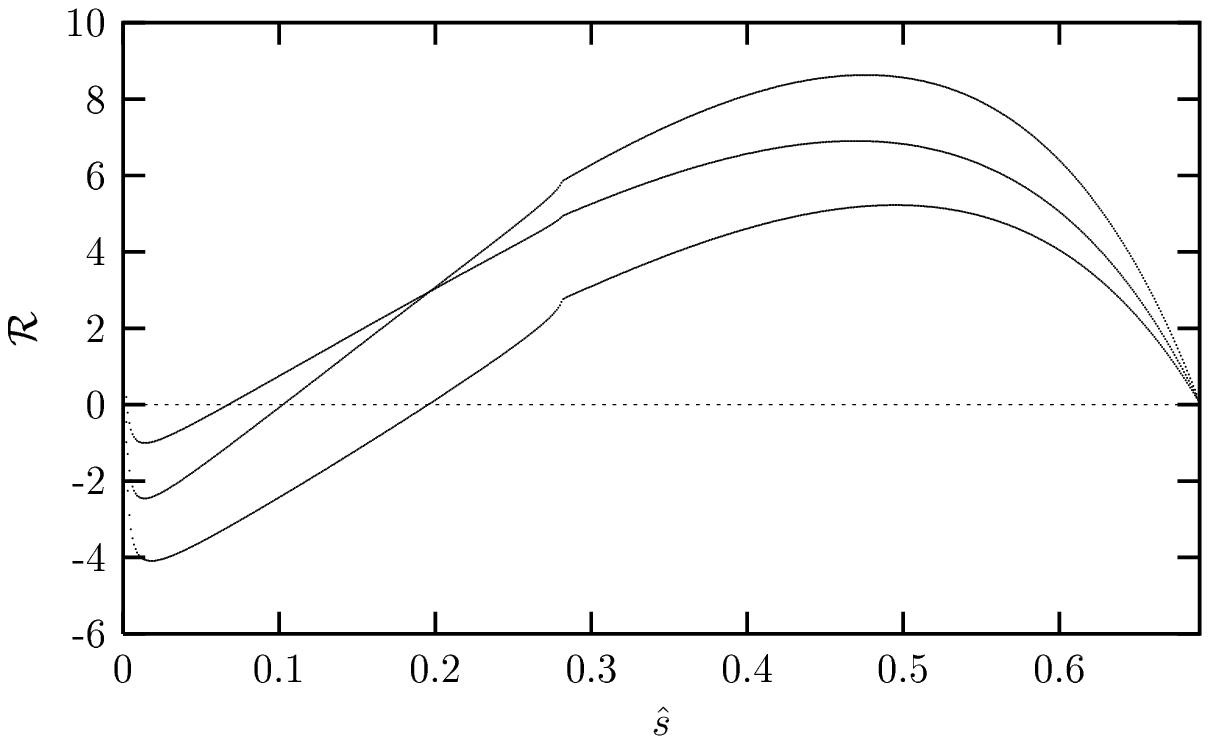}
\vskip -3.3truein
\caption[]{The dependence of ${\cal  R}$ on ${\hat s}$ 
which corresponds to the cases : $C_{LR}=C_{10}$ (bottom curve),
$C_{LR}=-C_{10}$ (middle
curve), and all the new coefficients are zero (top curve) .}
\label{fig2}
\end{figure}

Shown in Fig. 2 is  the dependence of ${\cal  R}$ on ${\hat s}$
when  $C_{LR}=-C_{10}$ (top curve), $C_{LR}=C_{10}$ (bottom curve),
and when all the new coefficients are zero (middle curve).
As we can see from the figure that, ${\cal  R}$ takes
the largest  value  when all the new coefficients are set to zero, for  
which case the zero position  occurs at $\hat{s}\sim 0.1$. 
Considering the non-zero values of the new operators, for instance,
when  $C_{LR}=-C_{10}$, with all the other new coefficients set equal to zero, 
${\cal R}$ crosses zero around ${\hat s} \sim 0.06$, 
and when  $C_{LR}=C_{10}$, 
with all the other new coefficients set equal to zero, 
it  crosses zero around ${\hat s} \sim 0.2$.
Namely, for  $C_{LR}=-C_{10}$, and  $C_{LR}=C_{10}$, 
the position of the zero shifts to left, and right, respectively, 
as compared to S.M, in accord with the new physics contributions to the
Wilson coefficients.  

In Fig. 3, we show the dependence of ${\cal  R}$ on ${\hat s}$
when  $C_{RR}=-C_{10}$ (top curve), $C_{RR}=C_{10}$ (bottom curve),
and when all the new coefficients are zero (middle curve). 
One notes that ${\cal  R}$  is less  sensitive to
the change in the values of $C_{RR}$, as compared to the first two cases (Fig. 1 and Fig. 2). That is,
both curves which correspond to  $C_{RR}=-C_{10}$, and $C_{RR}=C_{10}$
cases, behave similarly,  overlapping  up to $\hat s \sim 0.3$, and then
there is a gradual shift between these curves.
Therefore, they cross zero at the same value of ${\hat s}$ ($\sim 0.08$),
which is slightly different  from that of the S.M. 
Namely,  the position of the zero
gradually shifts to right, as compared to its S.M value.
Similar observations can be made for Fig. 4, when  $C_{RL}=-C_{10}$ (top curve), $C_{RL}=C_{10}$ (bottom curve),
and when all the new coefficients are zero (middle curve). One notes 
from the figure that the curves which correspond to $C_{RL}=-C_{10}$, and
$C_{RL}=C_{10}$ behave oppositely with respect to the  curve which
corresponds to S.M (middle curve), as compared to Fig. 3. Therefore,
unlike Fig. 3, the position of zero gradually shifts to left, as compared to the S.M.
\begin{figure}[htb]
\vskip -3.0truein
\centering
\epsfxsize=6in
\leavevmode\epsffile{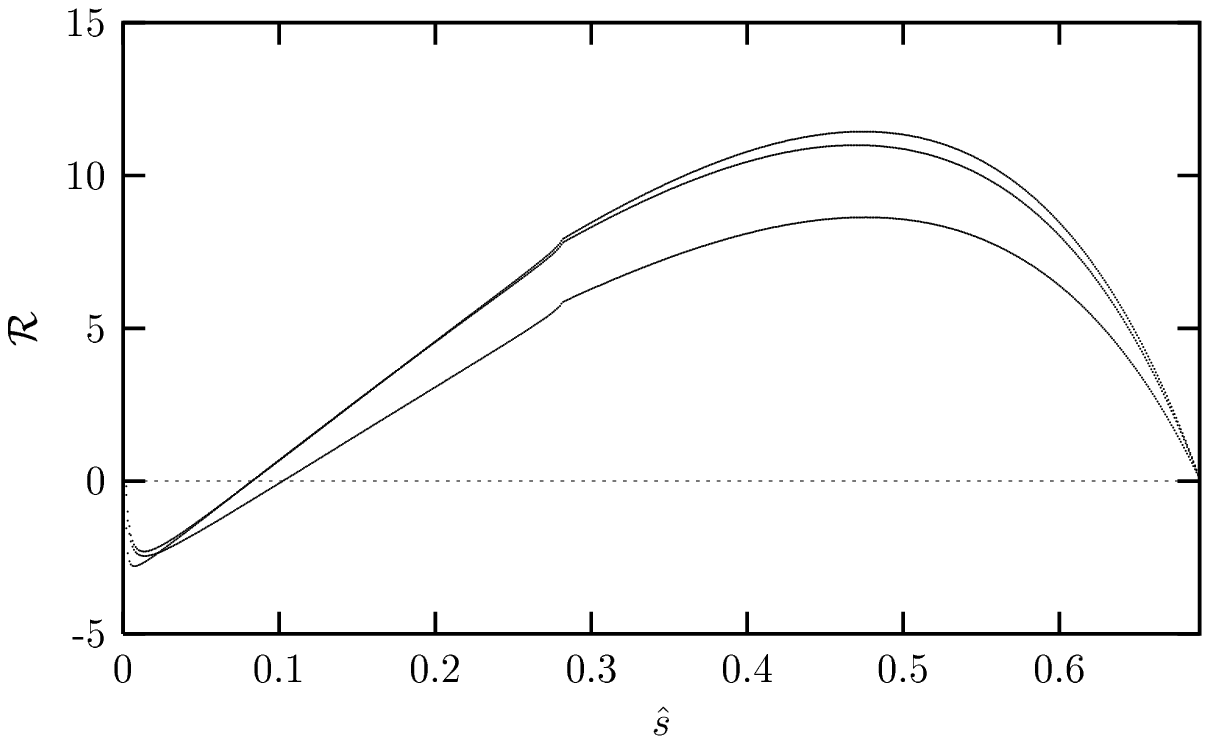}
\vskip -3.3truein
\caption[]{The dependence of ${\cal  R}$ on ${\hat s}$ 
which corresponds to the cases : $C_{RR}=-C_{10}$ (top curve),
$C_{RR}=C_{10}$ (middle
curve), and all the new coefficients are zero (bottom curve) .}
\label{fig3}
\end{figure}
\begin{figure}[htb]
\vskip -3.0truein
\centering
\epsfxsize=6in
\leavevmode\epsffile{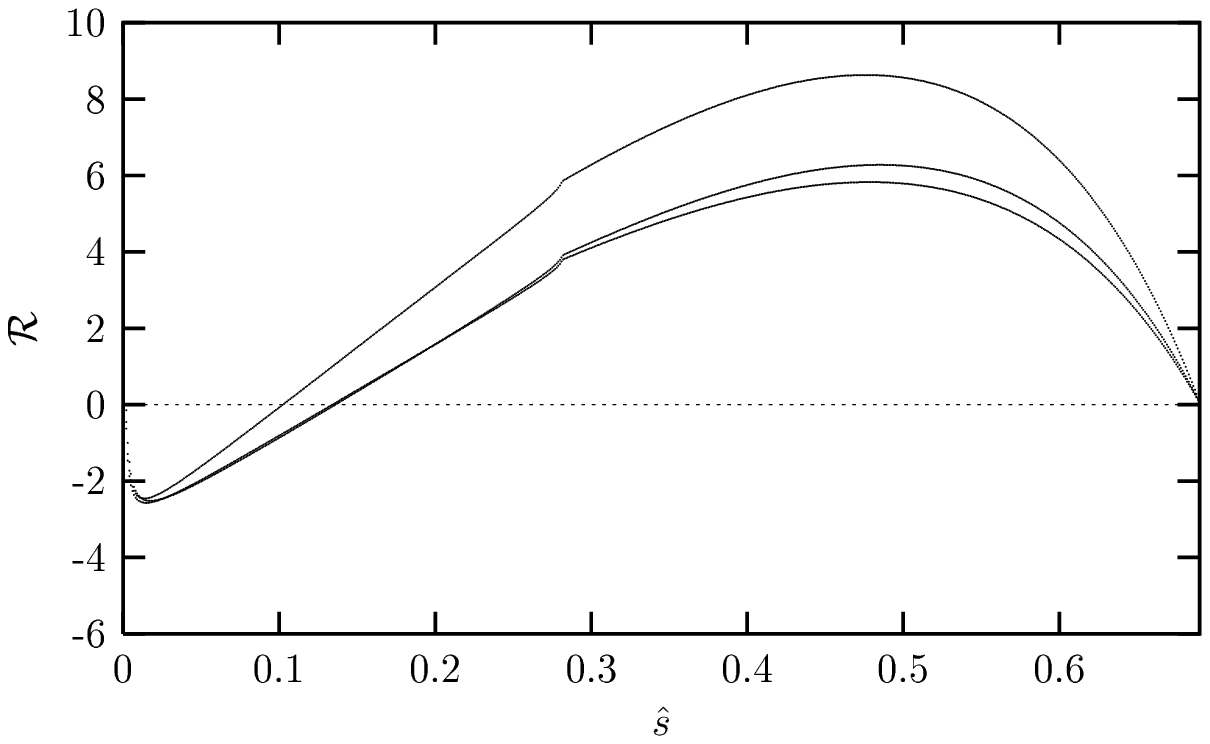}
\vskip -3.3truein
\caption[]{The dependence of ${\cal  R}$ on ${\hat s}$ 
which corresponds to the cases : $C_{RL}=C_{10}$ (middle curve),
$C_{RL}=-C_{10}$ (bottom
curve), and all the new coefficients are zero (top curve).}
\label{fig4}
\end{figure}

A comparative analysis of Figs 1-4 shows that
when $C_{LL(LR)}=C_{10}$, and all the other coefficients are zero,
the zero position shifts to left, and when  $C_{LL(LR)}=-C_{10}$  
it shifts to right, as compared to its value of S.M.
However, for  $C_{RR}=\pm C_{10}$, and $C_{RL}=\pm C_{10}$, with all the
other coefficients set to zero, the  shift in the 
zero position is to the right for the former, and to the 
left for the latter. One notes that although  the lepton mass effects are
included in our analysis, it can not give observable effects, as compared to
S.M,  since  the mass of $\mu$ ($m_\mu=0.105~\mbox{GeV}$) is quite small.
\begin{figure}[htb]
\vskip -3.0truein
\centering
\epsfxsize=6in
\leavevmode\epsffile{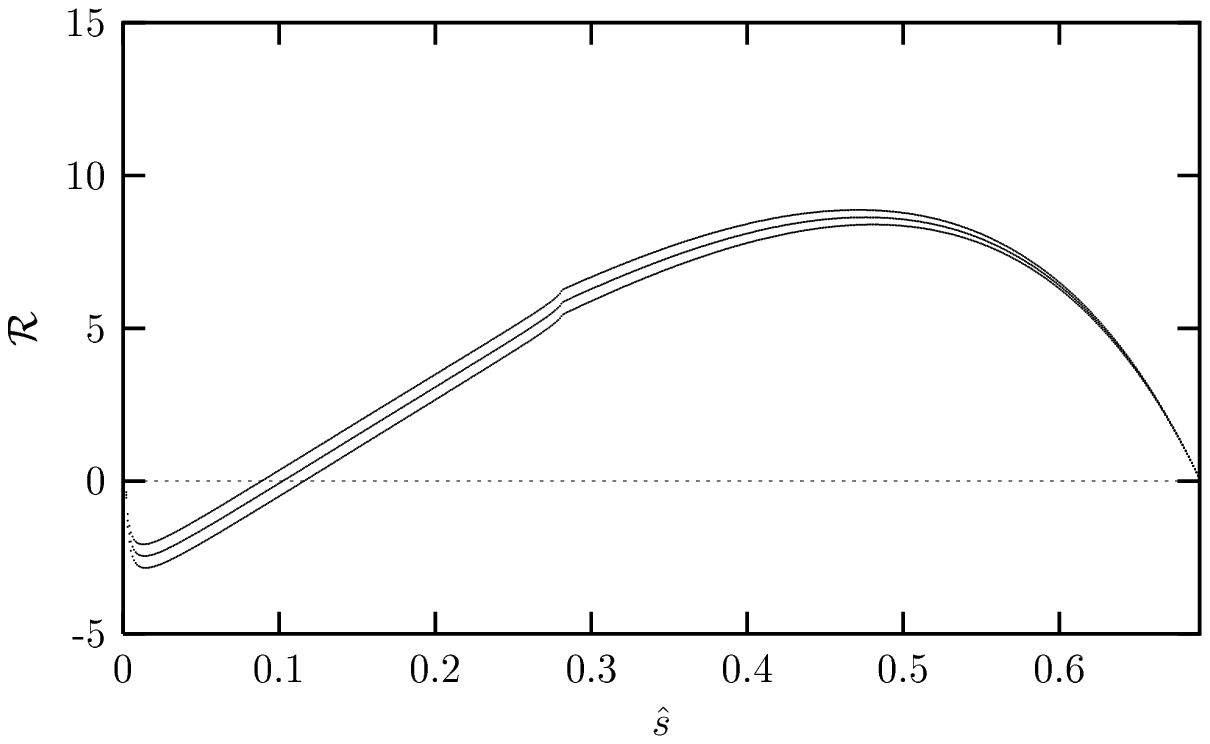}
\vskip -3.3truein
\caption[]{The dependence of ${\cal  R}$ on ${\hat s}$ 
which corresponds to the cases : $C_{LRLR}=C_{10}$ (top curve),
$C_{LRLR}=-C_{10}$ (bottom 
curve), and all the new coefficients are zero (middle curve) .}
\label{fig5}
\end{figure}

We show the dependence of ${\cal  R}$ on ${\hat s}$
on the scalar exchange operators, when  $C_{LRLR}=C_{10}$ (top curve),
$C_{LRLR}=-C_{10}$ (bottom
curve) in Fig. 5, and  $C_{LRRL}=C_{10}$ (top curve), $C_{LRRL}=-C_{10}$ (bottom curve)
in Fig. 6. In both figures, the middle curves 
correspond to the case when
all the new coefficients are equal to zero.
A comparative look at both figures suggests that,
when  $C_{LRLR(CLRRL)}=C_{10}$ (top curves of Fig. 5 and 6),  and
$C_{LRLR(LRRL)}=-C_{10}$ (bottom curves of Fig. 5 and 6), the dependence of
${\cal R}$ on the scalar exchange operators is exactly the same, 
and,  the positions of zeros shift to left
and right, as compared to S.M in both of these cases.
 \begin{figure}[htb]
\vskip -3.0truein
\centering
\epsfxsize=6in
\leavevmode\epsffile{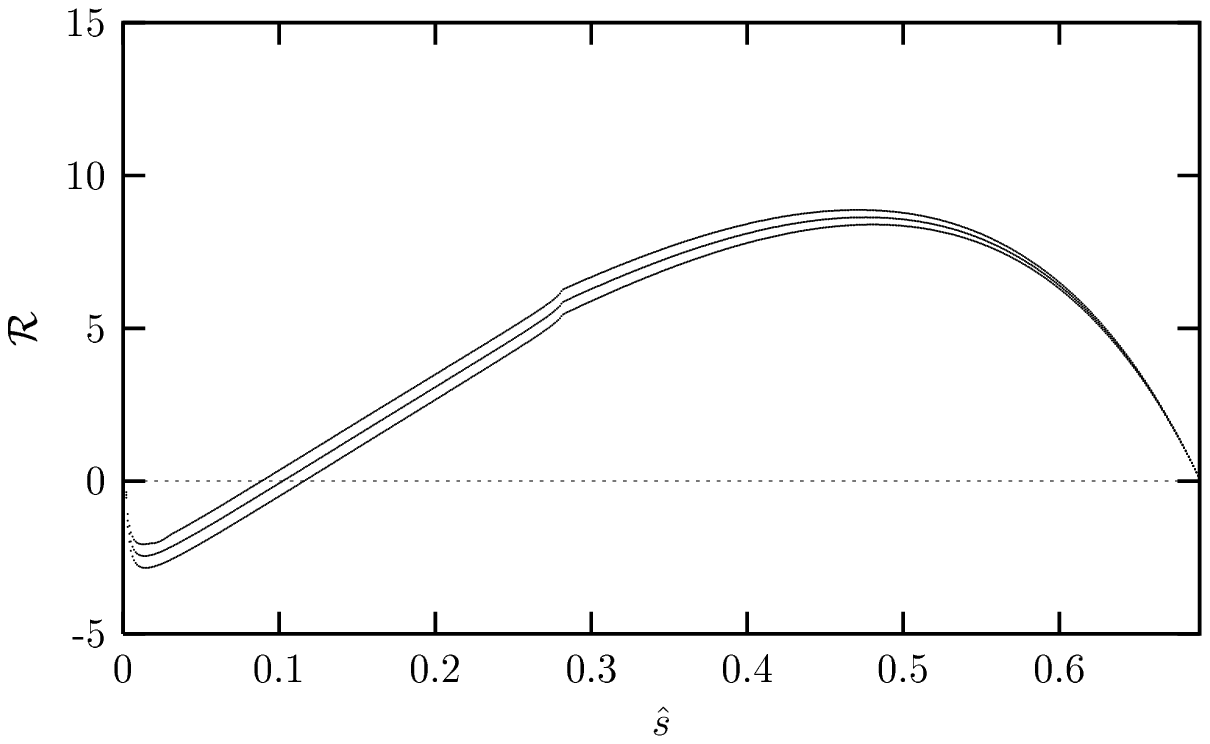}
\vskip -3.3truein
\caption[]{The dependence of ${\cal  R}$ on ${\hat s}$ 
which corresponds to the cases : $C_{LRRL}=C_{10}$ (top curve),
$C_{LRRL}=-C_{10}$ (bottom 
curve), and all the new coefficients are zero (middle curve). }
\label{fig6}
\end{figure}

\section{Conclusion}

In this work, we have studied the sensitivity of the zero position of the
forward backward asymmetry to the new physics effects.
It is found that the position of zero shifts in accord with the
new physics contributions to the Wilson coefficients.
Among all the new coefficients,
the zero of the asymmetry is highly sensitive to the sign and size of the 
vector--vector operators and opposite chirality counterparts
of the usual operators. The scalar--scalar four--fermion
operators, on the other hand, have mild effect on the 
zero of the asymmetry. 

Naturally, with increasing data and statistics, the experimental activities  in
B-physics  are expected to
give precise measurements in  semileptonic decays, and the
$\mbox{A}_{\small \mbox{FB}}$ is one of the key physical quantities that can be
measured. Therefore, it could  be appropriate to
provide an estimate about the number of events which is needed to measure
the forward-backward asymmetry with the  BaBar and Belle experiments.
$(i)$
Assuming that, BaBar will produce
$(3 -10) \times 10^{7}$ $b {\bar b}$ pairs in 1 year, for
${\cal L}=(3-10) \times  10^{33} cm^{-1} s^{-1}$ at
$\sqrt{s}=10~\mbox{GeV}$ \cite{lingel}, and 
$(ii)$
taking into account the experimentally relevant number of events required to
measure an asymmetry A of the decay with the 
branching ratio $Br(B \rar K^\ast \mu^+ \mu^-)=1.4 \times 10^{-6}$
\cite{aliev}
at the  $n \sigma$ level ($N=\frac{n^{2}}{Br A^{2}}$),
the number of events to observe $10\%$
$\mbox{A}_{\small \mbox{FB}}$  in the $B \rar K^\ast \mu^+ \mu^-$
decay (which is the average forward-backward asymmetry  
in the S.M) at $ 1 \sigma$ level can be estimated as $ 7 \times 10^{7}$.
Therefore, 1 year running of BaBar at  $ 1 \sigma$ level 
is sufficient  to measure the zero of the  $\mbox{A}_{\small \mbox{FB}}$, 
assuming that the asymmetry remains around  $10\%$.

The observation of the zero of the 
forward-backward asymmetry
(as suggested by the S.M) as well as possible shift
induced by the new physics effects both require the measurement
of $A_{FB}$ at different values of the dilepton invariant
mass. The estimate above is good for observing the
sign change in the asymmetry.
However, to observe the depletion of the asymmetry (asymmetry values
much smaller than $10\%$ ) requires much larger number
of $b{\bar b}$ pairs to be produced.

\noindent 

M. B would like to thank the Turkish Scientific and Technical Research
Council (T\"{U}B{\.I}TAK) for partial support under the project,
No:TBAG2002(100T108).

\end{document}